\documentclass[a4paper,10pt]{article}
\usepackage[dvips]{graphicx}

\title{Improving Robustness via Disjunctive Statements in Imperative Programming}

\author{Keehang Kwon, Sungwoo Hur, and Mi-Young Park\\
\textit{Dept. of Computer Engineering, DongA University}\\
\textit{Busan 604-714, South Korea}\\
khkwon, swhur, openmp@dau.ac.kr}

\newenvironment{numberedlist}
{\begin{list}{\makebox[20pt]{\hss(\arabic{itemno})\enspace}}
             {\usecounter{itemno}\labelwidth 20pt}}{\end{list}}

\newcounter{itemno}

\newcounter{itemno1}

\newcounter{itemno2}

\newcounter{exno}

\newcounter{defno}

\newenvironment{defn}{\refstepcounter{defno}\medskip \noindent {\bf
Definition \thedefno.\ }}{\medskip}

\newcommand{\sep}{\;\vert\;}

\newcommand{\oprove}{\vdash\kern-.6em\lower.7ex\hbox{$\scriptstyle O$}\,}

\newcommand{\Pscr}{{\cal P}}

\newcommand{\all}{\forall}

\newcommand{\ie}{{\em i.e.}}

\newsavebox{\lpartfig}
\newsavebox{\rpartfig}

\newenvironment{exmple}{
 \begingroup \begin{tabbing} \hspace{2em}\= \hspace{3em}\= \hspace{3em}\=
\hspace{3em}\= \hspace{3em}\= \hspace{3em}\= \kill}{
 \end{tabbing}\endgroup}

\newcommand{\lb}{\langle}
\newcommand{\rb}{\rangle}

     {\\* \hspace*{\fill} \end{trivlist}}

\newcommand{\seqor}{\cup}

\renewcommand{\seqor}{:}

\begin{document}
\maketitle
\begin{abstract}
To deal with failures as simply as possible,  we propose
a new foundation for the core (untyped) C,
which is based on a new logic called
task logic or imperative logic.
We then introduce a sequential-disjunctive statement of the form
$S \seqor R$.
This statement has the
following semantics: execute $S$ and $R$ sequentially.
 It is considered a success if at least one of $S, R$ is a
success.   This statement is useful for dealing with
inessential errors without explicitly catching them.
\end{abstract}

\section{Introduction}\label{sec:intro}

Imperative  programming is an important modern programming paradigm.
Successful languages in this paradigm includes C and Java.
 Despite much
attractiveness, imperative languages have
traditionally lacked fundamental notion of success/failure  for indicating
whether a statement can be successfully completed or not.
 Lacking such a notion, imperative programming relies on nonlogical, awkward devices such as exception handling
to deal with failures. One major problem with exception handling
is that  the resulting
language becomes complicated and not easy to use.

To deal with failures as simply as possible,  we propose
a new foundation for the core (untyped) C,
which is based on a new logic called
task logic \cite{Jap03,Jap08} or imperative logic.
The task logic expands the traditional t/f (true/false)
so as to include T/ F(success/failure).
The task logic interprets each statement
as T/F, depending on whether it can be successfully
completed or not. The premature exit of a statement due to failures can be problematic.
To avoid this, we  adopt ``all-or-nothing'' semantics discussed in
\cite{FF07} to guarantee
atomicity. Thus, if a failure occurs in the couse of executing a statement,
we assume that the machine rolls back partial updates.

We can then extend this ``logic-based''
C with other useful logical operations.  To improve robustness,
we introduce a sequential-disjunctive statement of the form
$S \seqor R$. Here, to avoid complications,
 we assume that $S$ and $R$ are independent of each other, \ie,
 no variables appear in  both $S$ and $R$.
This statement has the
following semantics: execute $S$ and $R$ sequentially.
 It is considered a success if at least one of $S, R$ is a
success.  This statement generates less exceptions, is easier to succeed,
and hence is more robust than other statements. This statement has
 the effect of  reducing the number of
exceptions to be dealt with without catching them.  It is useful for dealing with
inessential errors that can be ignored. For example,
the statement $S \seqor true$ has the effect of erasing all the possible exceptions raised in the course of executing
$S$ so that none of these exceptions can have  further interactions with the environment.

We also introduce a choice-disjunctive statement of the form $S\ else\ R$ which
is a logical version of  the  $try\ S\ catch\ R$ statement. This statement has the
following semantics: execute $S$.  If it is a success, then do nothing.
If it fails,  execute $R$.

The remainder of this paper is structured as follows. We describe the new language $C^L$
 in
the next section. In Section \ref{sec:modules}, we
present some examples.
Section~\ref{sec:conc} concludes the paper.

\section{The Language}\label{sec:logic}

The language is a subset of the core (untyped) C
 with some extensions. It is described
by $G$- and $D$-formulas given by the syntax rules below:
\begin{exmple}
\>$G ::=$ \>   $t\sep f \sep A \sep x = E \sep  G;G \sep   G\seqor G  \sep G\
 else\ G$ \\   \\
\>$D ::=$ \>  $ A = G\ \sep \all x\ D$\\
\end{exmple}
\noindent
In the rules above,
$A$  represents an atomic procedure definition of the form $p(t_1,\ldots,t_n)$.
A $D$-formula  is called a  procedure definition. $f$ denotes $false$ which correponds to
 a user-thrown exception.

In the transition system to be considered, $G$-formulas will function as the
main program (or statements), and a set of $D$-formulas enhanced with the
machine state (a set of variable-value bindings) will constitute  a program.

 We will  present an operational
semantics for this language via a proof theory. The rules  are formalized by means of what
it means to
execute the main task $G$ from a program $\Pscr$.
These rules in fact depend on the top-level
constructor in the expression,  a property known as
uniform provability\cite{MNPS91}. Below the notation $D;\Pscr$ denotes
$\{ D \} \cup \Pscr$ but with the $D$ formula being distinguished
(marked for backchaining). Note that execution  alternates between
two phases: the goal-reduction phase (one  without a distinguished clause)
and the backchaining phase (one with a distinguished clause).
The notation $S\ sand\ R$ denotes the following: execute $S$ and execute
$R$ sequentially. It is considered a success if both executions succeed.
The notation $not()$ denotes a failure.

\begin{defn}\label{def:semantics}
Let $G$ be a main task and let $\Pscr$ be a program.
Then the notion of   executing $\lb \Pscr,G\rb$ successfully and producing a new
program $\Pscr'$-- $ex(\Pscr,G,\Pscr')$ --
 is defined as follows:
\begin{numberedlist}

\item  $ex(\Pscr,t,\Pscr)$. \% True is always a success.

\item    $ex((A = G_1);\Pscr,A)$ if
 $ex(\Pscr, G_1)$ and  $ex(D;\Pscr, A)$.

\item    $ex(\all x D;\Pscr,A)$ if   $ex([t/x]D;
\Pscr, A)$. \% argument passing

\item    $ex(\Pscr,A)$ if   $D \in \Pscr$ and $ex(D;\Pscr, A)$. \% a procedure call

\item  $ex(\Pscr,x=E,\Pscr\uplus \{ \lb x,E' \rb \})$ if $eval(\Pscr,E,E')$.
\% $\uplus$ denotes a set union but $\lb x,V\rb$ in $\Pscr$ will be replaced by $\lb x,E' \rb$.

\item  $ex(\Pscr,G_1; G_2,\Pscr_2)$  if $ex(\Pscr,G_1,\Pscr_1)$  sand \\
  $ex(\Pscr_1,G_2,\Pscr_2)$.

\item  $ex(\Pscr,G_1\seqor G_2,\Pscr_2)$  if $ex(\Pscr,G_1,\Pscr_1)$  sand \\
  $ex(\Pscr,G_2,\Pscr_2)$. \% both $G_1$ and $G_2$ succeed.

\item  $ex(\Pscr,G_1\seqor G_2,\Pscr_2)$  if $not(ex(\Pscr,G_1,\Pscr_1))$  sand \\
  $ex(\Pscr,G_2,\Pscr_2)$. \% only $G_2$ succeeds.

\item  $ex(\Pscr,G_1\seqor G_2,\Pscr_1)$  if $ex(\Pscr,G_1,\Pscr_1)$  sand \\
  $not(ex(\Pscr,G_2,\Pscr_2))$. \% only $G_1$ succeeds.

\item $ex(\Pscr, G_1\ else\ G_2, \Pscr_1)$  if $ex(\Pscr,G_1,\Pscr_1)$

\item $ex(\Pscr, G_1\ else\ G_2, \Pscr_2)$  if $not(ex(\Pscr,G_1,\Pscr_1))$ sand \\
$ex(\Pscr,G_2,\Pscr_2))$.

\end{numberedlist}
\end{defn}

\noindent
If $ex(\Pscr,G,\Pscr_1)$ has no derivation, then the machine returns $F$, the failure.
For example, $ex(\Pscr,f,\Pscr_1)$ is a failure because it has no derivation.

\section{Examples}\label{sec:modules}

So far, we have considered only one kind of failures. In reality,
there are many kinds of failures in imperative programming. Thus, we need to expand
$f$ to include $f(e)$ for a user-thrown exception  $e$.
The notion of exception trees \cite{BM00} is then useful to organize failures, similar to
a file system in Unix and similar to an exception class in Java. Below we
assume that the machine returns an exception tree stored in $Failtree$ rather than just
$F$. We also assume that $/F$ is the root directory of $Failtree$ and
$/F/usr$ is the directory for user-thrown failures.
An exception can be derived from the parent exception. Exception trees
allow the programmer to select to deal with  failures at varying degrees
of specificity.
An example of the use of this construct is provided by the
following program which contains some basic file-handling rules.

\begin{exmple}
 $main$\\
$openfile();readfile()$\\
$else$ \\
\> $case\ Failtree\ of$\\
\>\> $/F/sys: \ldots$\\
\>\> $/F/usr/EOF: \ldots$;\\
$ x = factorial(4) $ \\
$readfile() =  (read( ) \not=  -1);\ldots\ else\ f(EOF)$
\end{exmple}

Our language  makes it possible to simplify the program if some statements are
inessential.  For example,
the following program explicitly tells the machine that the statement $openfile();readfile()$
is inessential and optional and thus it is OK not to perform the statement if it fails.

\begin{exmple}
 $main$\\
$(openfile();readfile()) \seqor $\\
$ x = factorial(4) $ \\
$readfile() =  (read( ) \not= -1);\ldots\ else\ f(EOF)$
\end{exmple}

\section{Conclusion}\label{sec:conc}

In this paper, we have considered an extension to the core C with
disjunctive statements. This extension allows statements of
the form  $S \seqor R$  where $S, R$ are statements.
These statements are
 particularly useful for dealing with inessential statements.

\section{Acknowledgements}

This work  was supported by Dong-A University Research Fund.

\bibliographystyle{ieicetr}

\begin{thebibliography}{1}



\bibitem{Jap03}
G.~Japaridze, ``Introduction to computability logic'', Annals  of Pure and
 Applied  Logic, vol.123, pp.1--99, 2003.

\bibitem{Jap08}
G.~Japaridze,   ``Sequential operators in computability logic'',
 Information and Computation, vol.206, No.12, pp.1443-1475, 2008.



\bibitem{FF07}
C.~Fetzer and P.~Felber, ``Improving Program Correctness with Atomic Exception
Handling'', Journal of Universal Computer Science, vol.13, no.8, pp.1047--1072,
2007.

\bibitem{BM00}
P.~Buhr and W.~Bok, ``Advanced Exception Handling Mechanisms'',
IEEE Transactions on Software Engineering, vol.26, no.9, pp.1--15, 2000.



\bibitem{MNPS91}
D.~Miller, G.~Nadathur, F.~Pfenning, and A.~Scedrov, ``Uniform proofs as a
  foundation for logic programming,'' Annals of Pure and Applied Logic, vol.51,
  pp.125--157, 1991.




\end{thebibliography}


\end{document}